\documentclass[12pt]{article}

\usepackage{subfigure}
\usepackage{graphicx}
\usepackage{url}

\newcommand{\aspas}[1]{{``#1''}}
\newcommand{\mcode}[1]{$\tt #1$}

\title{CodeCity for (and by) JavaScript}

\author{Marcos Viana, Andre Hora, Marco Tulio Valente}

\date{\small Department of Computer Science, \\ Federal University of Minas Gerais (UFMG), Belo Horizonte -- Brazil}

\begin{document}

\maketitle

\begin{abstract}
\noindent JavaScript is one of the most popular programming languages on the web. Despite the language popularity and the increasing size of JavaScript systems, there is currently a limited
number of visualization tools that can be used by developers to comprehend, maintain, and evolve JavaScript software. In this paper, we introduce JSCity, an implementation in JavaScript of the well-known Code City
software visualization metaphor. JSCity relies on JavaScript features and libraries to show ``software cities'' in standard web browsers, without requiring complex installation procedures.
We also report our experience on producing visualizations for 40 popular JavaScript systems using JScity.

\end{abstract}

\section{Introduction}
\label{introduction}

JavaScript is one of the most popular programming languages on the web. The language was designed in the middle of 1990 aiming to extend web pages with small pieces of executable code.  Since then, its popularity and relevance have only grown~\cite{about-time-to-take, about-time-to-take2}. Currently, JavaScript is used to develop complex applications in several domains, such as email clients, office applications, IDEs, etc. However, despite the increasing size and complexity of JavaScript systems, there is currently no visualization tools that can be used by developers to comprehend, maintain, and evolve JavaScript software.

In this short paper, we present JSCity, an open source tool that offers an intuitive way for representing, modeling and visualizing JavaScript source code using 3D visualizations.
Specifically, JSCity adapts to JavaScript the well-known {\em CodeCity} software visualization metaphor, originally proposed by Wettel and Lanza to visualize object-oriented code.
JSCity represents a JavaScript program as a city; folders are districts and files are sub-districts; the buildings are functions; inner functions are represented as buildings on the top of their enclosing function/building. The Number Of Lines of source Code (LOC) represents the height of the buildings/functions; the Number Of Variables (NOV) in a function correlates to the building's base size.
In the paper, we also report our experience on producing visualizations for 40 popular JavaScript systems using JScity.

This paper is organized as follows. 
Section \ref{metaphor} describes the city metaphor for the representation of code. Section~\ref{sec:tool} describes the implementation of JSCity in JavaScript.
Section \ref{results} illustrates the usage of JSCity in real JavaScript code.
Finally, related work is presented in Section \ref{related-work} and conclusions are provided in Section \ref{conclusion}.

\section{CodeCity for JavaScript}
\label{metaphor}

\begin{figure}[!ht]
  \centering
    \includegraphics[width=0.9\textwidth]{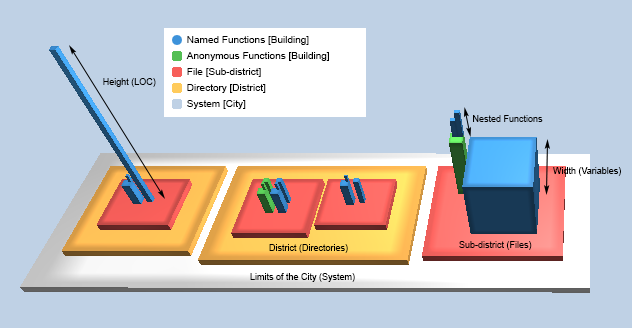}
  \caption{City metaphor for JavaScript.}
  \label{figure:metaphor}
\end{figure}

In the original code city metaphor, classes are represented as buildings. However, despite the possibility of emulating classes by means of prototypes~\cite{proceedings-of-the22}, JavaScript is not a class-based programming language, such as Smalltalk, Java, and C++. Therefore, we use buildings to represent JavaScript functions, as illustrated in Figure~\ref{figure:metaphor}.
The buildings' color is used to distinguish named functions (represented in blue) from anonymous functions (represented in green). The height of the buildings correlates with a function size, measured in number of lines of code; the width correlates with the number of variables of a function. Finally, directories are city districts and files are sub-districts. 
Districts and sub-districts are represented by the colors yellow and red, respectively. 

A common practice in JavaScript is to declare nested functions. To handle such functions, we adapt the city metaphor to display a building (nested function) on the top of another (the enclosing function).
The total height of a building is in fact the sum of the size of all nested functions (in lines of code).  The width of a parent function is the sum of its variables and the variables of the child functions, ensuring the construction of a wider building at the bottom and narrower buildings over it.  information.

\section{CodeCity by JavaScript}
\label{sec:tool}

JSCity is developed in JavaScript itself and uses the \textit{Esprima}\footnote{http://esprima.org} framework to generate an Abstract Syntax Tree (AST) for the target code. The \textit{ThreeJS} framework\footnote{https://threejs.org} is used to design the 3D city. ThreeJS offers several graphical resources that enable scenes representation in 3D. 
Among the offered features, we enumerate the creation of scenes, geometries, animations, and perspectives without page reloading. 
These features are used in our tool to draw the city, to make it possible to navigate by city elements, to perform \textit{zoom} operations, and to change the camera angle and cursor position with function.

\begin{figure}[ht!]
  \centering
    \includegraphics[width=0.9\textwidth]{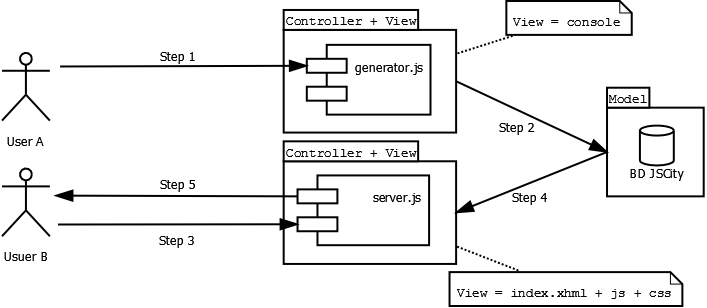}
  \caption{Component diagram of JSCity tool.}
  \label{figure:js-city-model}
\end{figure}

Figure~\ref{figure:js-city-model} presents the steps that are needed to produce a code city visualization. First, during an offline phase, the \textit{Esprima} parser is called to generate an AST. After that,  the data required for producing a city visualization is persisted in a database. During an online phase, a {\em City Drawer} component is responsible for retrieving the data required for producing a city visualization and to draw the city using the \textit{ThreeJS} framework.

\section{Examples}
\label{results}

We use JSCity to produce code cities for 40 well-known JavaScript systems, including popular libraries and frameworks, code editors, browser plugins, and games. 
The complete list of systems, with the respective cities, is available at: \url{https://github.com/ASERG-UFMG/JSCity/wiki/JSCITY}. 

In this section, we present and discuss the code cities of three systems:

\begin{itemize}

\item Figures~\ref{figure:angular-i18n} and \ref{figure:angular-core} present the city of {\sc AngularJS}, a popular MVC-based framework. The system has 233,785 lines of source code, 20 directories, 863 files, 10,362 anonymous functions, and 6,050 named functions. We can see that {\sc AngularJS}'s city has two main areas: a district with small buildings (Figure~\ref{figure:angular-i18n}) and an area with many skyscrapers (Figure~\ref{figure:angular-core}). In fact, the directory (district) in Figure~\ref{figure:angular-i18n} has many small anonymous functions (in green) for internationalization, which are located in directory \mcode{ngLocale}. By contrast, the district in Figure~\ref{figure:angular-core} includes functions responsible for the system core. The tallest building is function \mcode{\$compile}. This function \aspas{compiles an HTML string or DOM into a template and produces a template function}.

\begin{figure}[!h]
\centering
 \centering
\includegraphics[width=.5\linewidth]{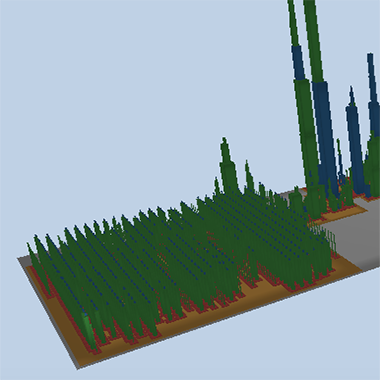}
\caption{AngularJS: Internationalization functions}
\label{figure:angular-i18n}
\end{figure}

\begin{figure}[!h]
  \centering
  \includegraphics[width=.5\linewidth]{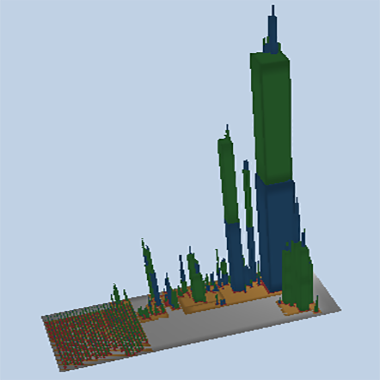}
  \caption{AngularJS:Core Functions}
  \label{figure:angular-core}
\end{figure}

\item Figure~\ref{figure:jquery} presents the city of {\sc jQuery}, the popular DOM manipulation library for JavaScript. 
The code has mostly anonymous functions (green buildings). The big buildings represent core event-handling functions, like an anonymous function in \mcode{jquery/ajax.js} with 798 LOC.
We also see through the visualization that directories are organized in large modules, such as \mcode{core}, \mcode{event}, and \mcode{data}.

\begin{figure}[ht!]
  \centering
    \includegraphics[width=0.6\textwidth]{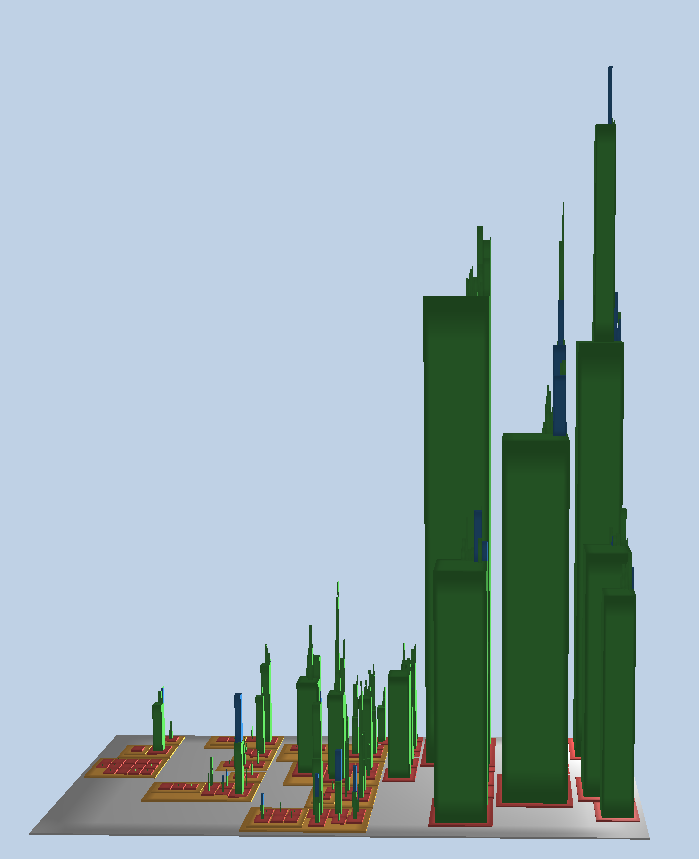}
  \caption{jQuery city}
  \label{figure:jquery}
\end{figure}

\item Figure~\ref{figure:bower} presents the city of Bower, a popular package management system for JavaScript.
We observe that most of functions are anonymous (green buildings), but named functions are also common (blue buildings). 
Besides that, the use of nested functions is very common (buildings on top of other buildings).

\begin{figure}[ht!]
  \centering
    \includegraphics[width=0.7\textwidth]{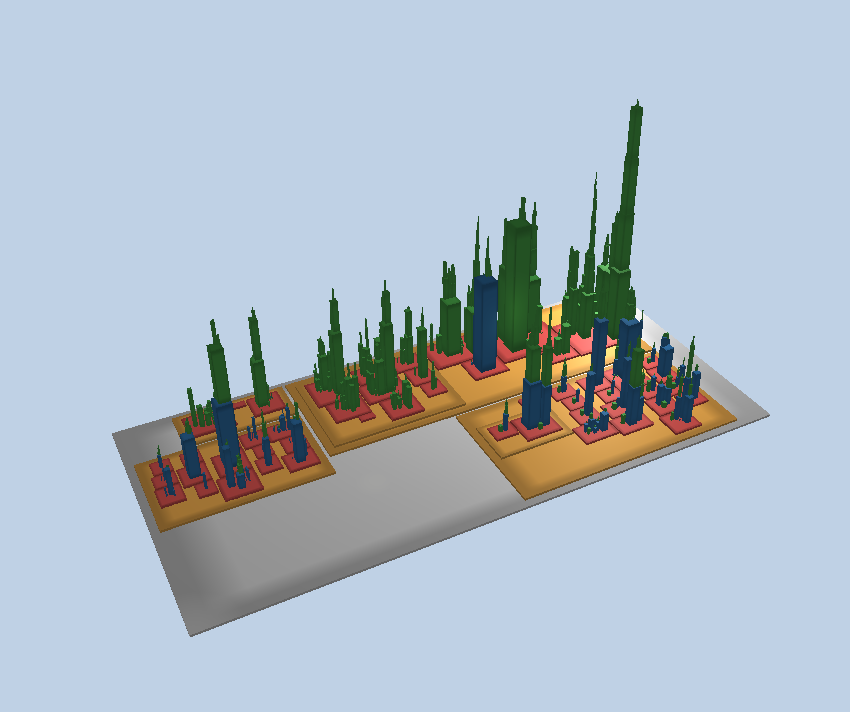}
  \caption{Bower city}
  \label{figure:bower}
\end{figure}

\end{itemize}

From our selection of systems, it is possible to infer some patterns of development, summarized as follows. Tall and wide buildings are usually core functions of the systems;
large districts with small and regular layouts represent, for example, internationalization features. Green buildings (anonymous functions) are largely used in practice as
well as buildings on top of other buildings (nested functions).

\section{Related Work}
\label{related-work}

In this work, we presented JSCity, which is inspired by CodeCity~\cite{3d-visualization,visualization-software}. 
CodeCity aims to make software analysis, where systems are shown as navigable and interactive 3D cities. Classes are represented as city buildings, while modules are represented as districts. 
Number of methods is the height of the buildings, the number of variables represents the width, and the number of lines of code is represented by colors -- from dark gray (lowest number) to intense blue (greater number). 
CodeCity is available in Eclipse~\cite{codecity_eclipse} and Moose tool~\cite{codecity_moose}. 
However, such cases are restricted to the analysis of the object-oriented languages Java and Smalltalk.
Finally, other visualization---not necessarily in the form of cities---have already been proposed for other system dimensions, such as bug analysis~\cite{hora12}.

\section{Conclusion}
\label{conclusion}

JSCity extends CodeCity for JavaScript and provides---to our knowledge---the first source code visualization tool for this language.
The city metaphor was adapted to represent JavaScript systems, for instance, by supporting the visualization of anonymous and nested functions. 
Moroever, JSCity is easily accessible because it runs directly in a web page, without requiring the installation of any visualization-specific software
As future work, we plan to survey developers about posssible aplications and use of code cities. We also plan to provide support to other languages, but keeping the visualization engine implemented in JavaScript.

\section*{Acknowledgments}

This research was supported by CNPq and FAPEMIG.

\bibliographystyle{plain}
\bibliography{jscity}

\end{document}